\documentclass[aps,
amsmath,amssymb,
preprint,
superscriptaddress
]{revtex4-1}

\usepackage[whole]{bxcjkjatype} 
\usepackage[unicode]{hyperref} 

\usepackage{graphicx}
\usepackage{color}
\usepackage{amsmath}
\usepackage{amssymb}
\usepackage{physics}
\usepackage{amsfonts}
\usepackage{colortbl}
\usepackage{xcolor}
\usepackage{dcolumn}
\usepackage{bm}

\usepackage[utf8]{inputenc}
\usepackage[T1]{fontenc}
\usepackage{mathptmx}
\usepackage{etoolbox}

\usepackage{lineno}

\makeatletter
\def\@email#1#2{%
 \endgroup
 \patchcmd{\titleblock@produce}
  {\frontmatter@RRAPformat}
  {\frontmatter@RRAPformat{\produce@RRAP{*#1\href{mailto:#2}{#2}}}\frontmatter@RRAPformat}
  {}{}
}%
\makeatother
\begin{document}


\title{Enhancement of $J \times B$ electron acceleration with the micro-structured target and picosecond high-contrast relativistic-intensity laser pulse}
\author{Ryunosuke Takizawa}
 \email{takizawa.ryunosuke.ile@osaka-u.ac.jp}
 \affiliation{Institute of Laser Engineering, The University of Osaka, 2-6 Yamada-Oka, Suita, Osaka 565-0871, Japan}
\author{Yuga Karaki}
 \affiliation{Department of Physics, Graduate School of Science, The University of Osaka, Toyonaka, 560-0043, Japan}
 \affiliation{Institute of Laser Engineering, The University of Osaka, 2-6 Yamada-Oka, Suita, Osaka 565-0871, Japan}
\author{Hiroki Matsubara}
 \affiliation{Department of Physics, Graduate School of Science, The University of Osaka, Toyonaka, 560-0043, Japan}
 \affiliation{Institute of Laser Engineering, The University of Osaka, 2-6 Yamada-Oka, Suita, Osaka 565-0871, Japan}
\author{Rinya Akematsu}
 \affiliation{Department of Physics, Graduate School of Science, The University of Osaka, Toyonaka, 560-0043, Japan}
 \affiliation{Institute of Laser Engineering, The University of Osaka, 2-6 Yamada-Oka, Suita, Osaka 565-0871, Japan}
\author{Ryou Oomura}
 \affiliation{Department of Physics, Graduate School of Science, The University of Osaka, Toyonaka, 560-0043, Japan}
 \affiliation{Institute of Laser Engineering, The University of Osaka, 2-6 Yamada-Oka, Suita, Osaka 565-0871, Japan}
\author{Law King Fai Farley}
 \affiliation{Institute of Laser Engineering, The University of Osaka, 2-6 Yamada-Oka, Suita, Osaka 565-0871, Japan}
\author{Hiroshi Azechi}
 \affiliation{Institute of Laser Engineering, The University of Osaka, 2-6 Yamada-Oka, Suita, Osaka 565-0871, Japan}
\author{Natsumi Iwata}
 \affiliation{Institute of Laser Engineering, The University of Osaka, 2-6 Yamada-Oka, Suita, Osaka 565-0871, Japan}
\author{Tomoyuki Johzaki}
 \affiliation{Graduate School of Advanced Science and Engineering, Hiroshima University, 1-4-1 Kagamiyama, Higashi-Hiroshima, Hiroshima, 739-8511, Japan}
 \affiliation{Institute of Laser Engineering, The University of Osaka, 2-6 Yamada-Oka, Suita, Osaka 565-0871, Japan}
\author{Yasuhiko Sentoku}
 \affiliation{Institute of Laser Engineering, The University of Osaka, 2-6 Yamada-Oka, Suita, Osaka 565-0871, Japan}
\author{Shinsuke Fujioka}%
 \affiliation{Institute of Laser Engineering, The University of Osaka, 2-6 Yamada-Oka, Suita, Osaka 565-0871, Japan}
  \affiliation{National Institute for Fusion Science, 322-6 Oroshi, Toki, Gifu 509-5292, Japan}


\date{\today}

\begin{abstract}
Efficient generation of multi-hundred-keV electrons is essential for isochoric heating and can influence ion acceleration. 
We investigated electron acceleration from copper-oleate foil targets—either planar or coated with a gold mesh structure (bar width 5 $\mu$m, spacing 7.5 $\mu$m, thickness $\sim$6 $\mu$m)—irradiated by 1.5 ps, 350 J LFEX laser pulses. 
Two laser contrast conditions were examined: high ($\sim 10^{10}$, with plasma mirror) and low ($\sim 10^{8}$, without plasma mirror). 
Using Cu–K$_\alpha$ emission mapping, we found that under high-contrast irradiation, the micro-structured target enhanced the laser-to-electron conversion efficiency from 4.9\% to 14\%, attributed to multiple internal reflections that enhance $J \times B$ acceleration. 
In contrast, under low-contrast conditions, the structures were filled with pre-plasma before the main pulse, and no enhancement was observed. 
These results demonstrate that both fine-scale structuring and high contrast are crucial for maximizing $J \times B$-driven electron generation in laser–plasma interactions. 
Our findings suggest a practical approach to improving laser–plasma coupling efficiency by exploiting micro-structured surfaces and contrast-controlled irradiation.
\end{abstract}

\pacs{}

\maketitle 

\section{Introduction}

Particle acceleration driven by relativistically intense lasers \cite{Esarey2009-lj} is indispensable 
for a broad spectrum of applications, including isochoric plasma heating for inertial confinement fusion \cite{Tabak1994-mw, Kodama2001-mg, Jarrott2016-cb, Sakata2018-fp, Matsuo2020-jg}, compact ion sources \cite{Macchi2013-pu, Wilks2001-mb, Habara2003-ae, Yogo2009-vn}, and ultrafast proton radiography for high-energy-density physics \cite{Schaeffer2023-kw, Nishiuchi2015-yf, Gao2016-sh}. 
Among these diverse applications, enhancing the efficiency of laser-driven particle acceleration has become particularly urgent for emerging technologies such as nuclear-fusion power generation.
The optimum particle energy and beam directivity depend on the specific application, so the underlying acceleration mechanism must be selectively enhanced.
Electrons are accelerated by various mechanisms whose characteristics depend on laser parameters and plasma scale length. 
When the plasma scale length $L$ greatly exceeds the laser wavelength $\lambda_{\rm L}$ ($L \gg \lambda_{\rm L}$), direct laser acceleration (DLA) \cite{Hartemann1995-jl} and laser‑wakefield acceleration (LWFA) \cite{Tajima1979-sb} dominate in underdense plasmas, while the resonant absorption \cite{Forslund1975-op} and vacuum heating \cite{Brunel1987-jd} are preferentially driven at oblique laser incidence when the laser light interacts with the overcritical plasmas.
The ubiquitous mechanism is $J \times B$ acceleration \cite{Kruer1985-dy,Wilks1992-yp}, in which the mean electron energy follows ponderomotive scaling. 
The aforementioned acceleration mechanisms recur when the laser pulse duration exceeds a few ps, giving rise to stochastic acceleration \cite{Sentoku2002-fo, Sorokovikova2016-ec, Kojima2019-zy, Iwata2019, Shen2022}, for which the high-energy tail of the electron spectrum can be fitted by a power-law distribution \cite{Fermi1949}.
The laser absorption fraction is highly sensitive to the plasma scale length: it deteriorates if the scale length is either too long \cite{Ivanov2017-rl} or fully plasma‑free \cite{Ping2008-sl}. 

High-contrast lasers are considered promising for controlling the properties of accelerated electrons. 
This is because, in contrast to low-contrast pulses, high-contrast lasers prevent the target from being prematurely filled with pre-plasma, allowing the target structure to be preserved until the arrival of the main pulse. 
As a result, laser–plasma interactions can be effectively controlled by the target geometry. 
Cone‑shaped targets exemplify this "geometric control" strategy; they channel high‑contrast lasers toward the tip and have already demonstrated electron‑beam collimation \cite{Takizawa2025-ry}.
Structured targets with nano- to micro-scale features on their surface have been proposed to increase laser absorption efficiency \cite{Habara2016-ss, Bailly-Grandvaux2020-tw, Tanaka2023-xz, Rocca2024-yy}. 
Two principal effects arise from these structures: (i) filling the inter-structural gaps with plasma produces a large-scale under-dense plasma, and (ii) multiple internal reflections of the laser increase the number of interaction cycles. 
If the pulse duration exceeds a few ps, these structures will be destroyed before the peak of the main pulse arrives, making the benefits less pronounced than those expected with a femtosecond laser pulse.

Here, we have demonstrated that a structured target increases the number of $J \times B$ accelerated electrons even for a picosecond laser pulse when the pre-formed plasma was suppressed, "{\it pre-plasma free}" condition. 
The pre-plasma free condition is realized by introducing a plasma mirror (PM) to enhance the laser contrast.
Additionally, we implemented a diagnostic method that allows for a more detailed characterization of the accelerated electrons.
Previous studies rely mainly on particle spectrometry; however, it only detects the escaping electrons whose energies in the target exceed the sheath-field potential excited around an irradiated target and also exceed the stopping power of the plasma.
The other schemes with characteristic X‑ray diagnostics utilized multilayer tracers \cite{Vaisseau2017-nb, Li2020-wj} or one‑dimensional wire tracers \cite{Vauzour2012-be, Ma2012-pa}, which prevent simultaneous measurement of the spatial and energy distributions of accelerated electrons. 
We have achieved two‑dimensional mapping of electron propagation in solid targets, allowing direct determination of both the laser‑to‑electron conversion efficiency and the divergence angle of $J \times B$ accelerated electrons.

Sections II and III detail the experimental setup and analysis methodology; Section IV summarizes the results; Section V discusses the conclusions and prospects.


\section{Experimental Configurations}
\label{sec:exp_conf}

Experiments were performed with the four beams of LFEX laser (wavelength: 1.053 $\mu$m, pulse: 1.5 ps, FWHM spot: $(130 \pm 30)\ \mu{\rm m}$) at The University of Osaka.
A PM was used to enhance the pulse contrast.
The PM is a spherical glass substrate with an anti-reflection coating; when the incident fluence exceeds the threshold (typically $6 \mathrm{-} 7\ \mathrm{J/m^{2}}$) \cite{Doumy2004-ag}, a plasma is formed on the surface, reflecting the main pulse, whereas pre-pulse components below the threshold are transmitted without reflection.
The reflectivity of the PM ($\sim$ 60\%) was obtained from previous research on LFEX \cite{Arikawa2016-so}.
Two contrast conditions were investigated: a \emph{high-contrast} case with the PM and a \emph{low-contrast} case without the PM.
The laser contrast was $10^{10}$ in the high-contrast case and $10^{8}$ in the low-contrast case at 100 ps before the main pulse.

Experiments were carried out with two target configurations: a structured target and a planar (flat-foil) target.
Each target employed a tracer consisting of a $200\ \mu$m-thick layer of copper oleate ($9.7\%$ Cu by weight, $\rho = 0.97\ \mathrm{g/cm}^{3}$).
The characteristics of $J \times B$ accelerated electrons were diagnosed by measuring both the absolute yield and the spatial distribution of Cu–K$_\alpha$ emission.
The characteristic X-rays are emitted when electrons collide with Cu atoms in the tracer with a known probability.
A 5 $\mu$m-thick gold foil was laminated onto the front surface of the tracer.
In the structured target, an additional gold mesh ($6 \pm 1\ \mu$m thick; bar width $5\ \mu$m; space width $7.5\ \mu$m) was bonded to the gold foil to form the surface structure.
A $1\ \mathrm{mm} \times 1\ \mathrm{mm}$ Ta block was attached to the rear surface of every target to suppress electron refluxing.

\begin{figure}[h]
  \centering
  \includegraphics[width=0.8\linewidth]{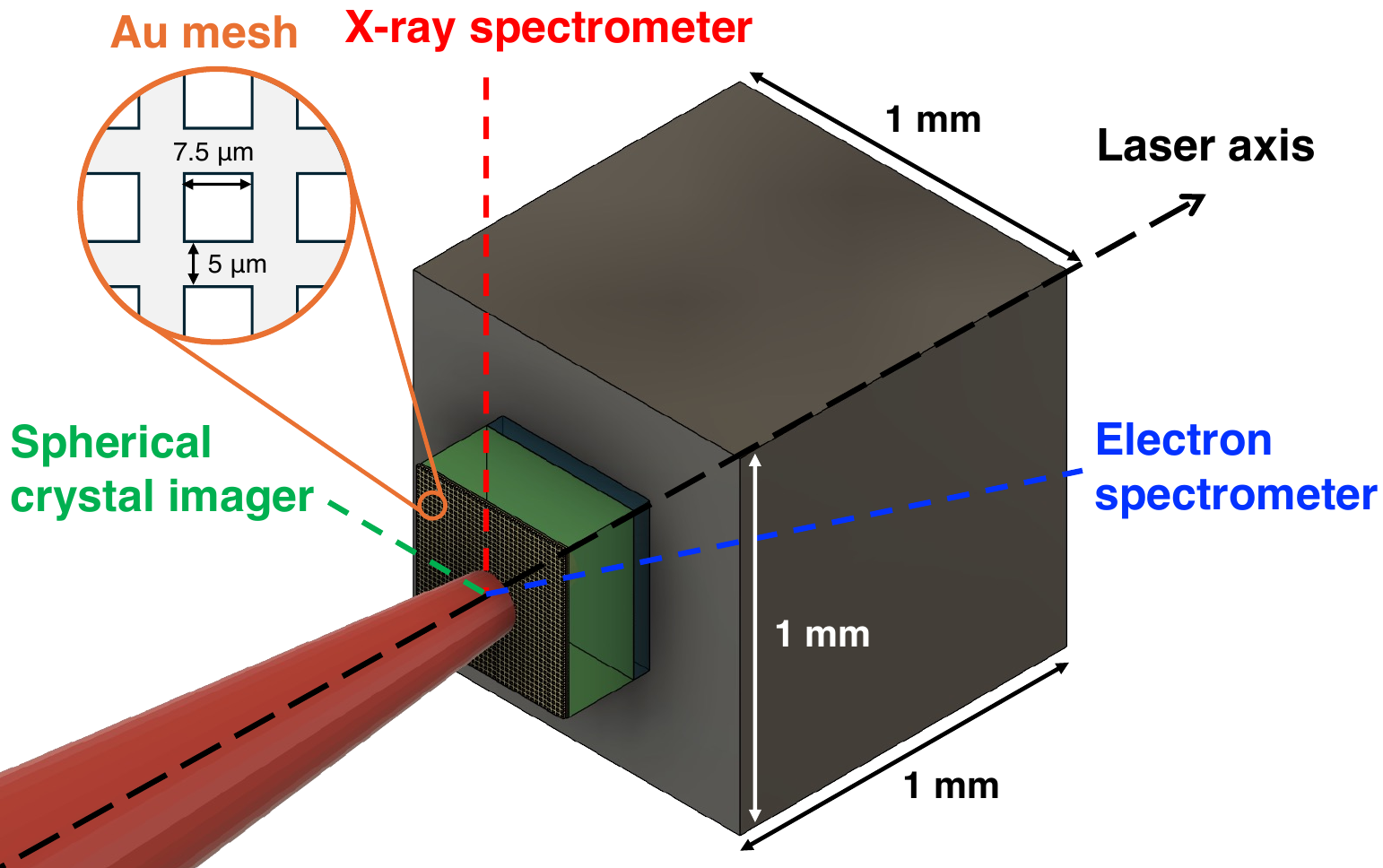}
  \caption{
  {\bf Experimental configuration}
  The gold mesh was attached  to construct the surface structure.
  Copper oleate with a thickness of 200 $\mu$m was used as the tracer.
  Electron refluxing was suppressed by the high-volume 1 mm square Ta.
  The X-ray spectrometer and spherical crystal imager were allocated along the vertical axis of the laser.}
  \label{fig:target}
\end{figure}

The absolute Cu–K$_\alpha$ photon yield was measured with an absolutely-calibrated flat highly-oriented-pyrolytic-graphite (HOPG) Bragg spectrometer, while the spatial distribution of Cu–K$_\alpha$ emission was obtained with a spherical crystal imager (SCI).
Both diagnostics were aligned perpendicular to the laser axis.
Electrons above $2\ \mathrm{MeV}$ were measured by an electron spectrometer (ESM) located on the Ta-block side of the target.
The lower limit of the electron energy, approximately 2 MeV, is determined from the continuous slowing down approximation (CSDA) range of the Ta block ($1.7\ {\rm g/cm}^{2}$) \cite{NIST124}.
Figure \ref{fig:target} shows the experimental configuration.

\section{Reconstruction method of Cu–K$_{\alpha}$ emission}
\label{sec:sim}

\begin{figure}[h]
  \centering
  \includegraphics[width=0.8\linewidth]{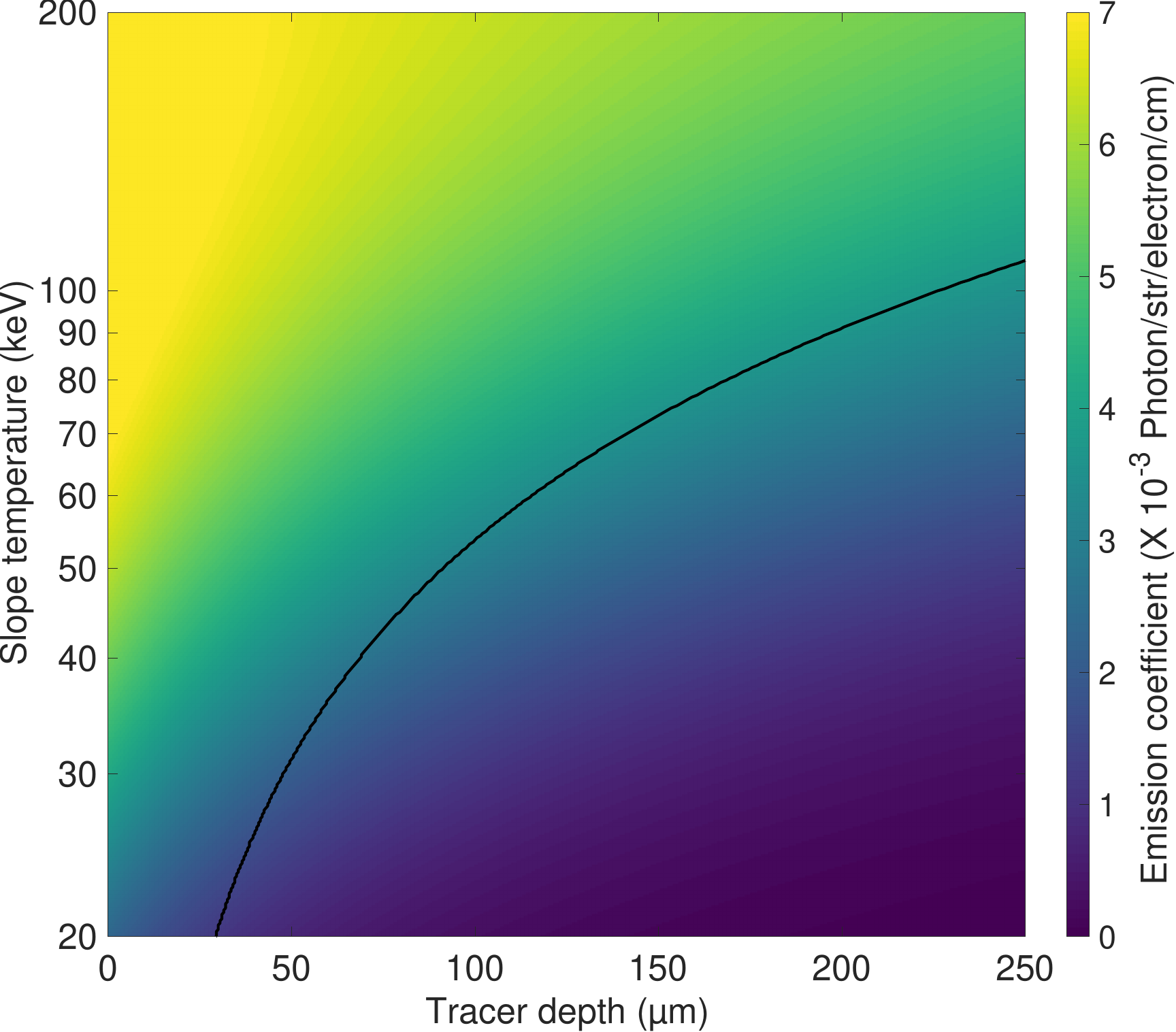}
  \caption{
  {\bf The mapping of the emission coefficient per electron}
  A Boltzmann distribution was assumed for calculating the map.
  The black line marks the 50\% contour of the emission coefficient at the tracer surface.
  The tracer emits uniformly in space at electron temperatures above about 200 keV.
  }
  \label{fig:emi_dens}
\end{figure}

We assumed cylindrical symmetry about the laser axis and calculated the three-dimensional spatial distribution of Cu–K$_\alpha$ emission. 
The simulated emission profile was subsequently integrated along the detector line of sight and compared with the experimental data.
The spatiotemporal laser-intensity profile is
\begin{equation}
    I\qty(r, t) = I_0 \exp(-\frac{r^2}{2\sigma_r^2} -\frac{t^2}{2\sigma_t^2}),
\end{equation}
where $r$ is the radial coordinate, $t$ is time, $\sigma_r$ is the variance of the laser spot, $\sigma_t$ is the variance of the pulse duration, $E_{\rm L}$ is the laser energy, and $I_0 = E_{\rm L}/\bigl[(2\pi)^{3/2}\sigma_r^2\sigma_t\bigr]$ is the peak laser intensity.
The Boltzmann distribution was used for the energy distribution of electrons which mean energy follows the ponderomotive scaling,
\begin{equation}
    T_\mathrm{e}(r,t)=
  m_e c^{2}\qty[
  \sqrt{1+\frac{I(r,t)\,\lambda^{2}_{\rm L}}
               {1.37\times10^{18}\,\mathrm{W\,cm^{-2}\,\mu m^{2}}}}
  -1
  ],
\end{equation}
where $m_e$ is the electron mass and $c$ is the speed of light.
The emission coefficient relative to electron propagation can be calculated by combining the emission probability and the electronic stopping power of the target as shown below.
First, the electron energy distribution function $f\qty(E,z,T_e)$ which depends on the initial slope temperature ($T_e$),  evolves in space as follows:
\begin{align}
f\qty(E,z,T_e) = f\qty[g(E,z), 0, T_e]\abs{\dv{E}g(E,z)}, \\
g(E,z) = E + \int^{z}_{0}S\qty[g(E,z^\prime)]\dd{z^\prime},
\end{align}
where $E$ is the electron kinetic energy, $z$ is the propagation distance, $T_e$ is the initial mean electron energy, and $S(E)$ is the stopping power.
The emission coefficient per electron after traversing a depth $z$ is expressed as
\begin{align}
    j_{\rm 1D}\qty(z,\,T_\mathrm{e}) = 
    \frac{\int_{\rm 10\ keV}^{\infty} v\qty(E)f\qty(E,z,T_e)n_{\rm Cu}\sigma_{\rm K_\alpha}/4\pi\dd{E}}
    {\int_{0}^{\infty} v\qty(E)f\qty(E,0,T_e)\dd{E}}
\end{align}
where $v\qty(E)$ is the electron velocity, $n_{\rm Cu}$ is the number density of Cu ions, and $\sigma_{\rm K_\alpha}$ is the electron-impact K-shell ionisation cross-section.
Assuming a Boltzmann distribution $f(E,0)=\exp(-E/T_e)$ at the target surface, the calculated $j_{\rm 1D}\qty(z,T_\mathrm{e})$ mapping is shown in Fig.~\ref{fig:emi_dens}.
Cu–K$_\alpha$ photons are emitted from deeper regions as the electron energy increases, and the entire 200 $\mu$m thick tracer begins to uniformly emit once $T_e$ exceeds approximately 200 keV.
We calculated the emission distribution for a copper-oleate target.
The Hombourger \cite{Hombourger1998-rc} and Davies \cite{Davies2013-wm} models were used for $\sigma_{\rm K_\alpha}$, while stopping powers were taken from the NIST ESTAR database \cite{NIST124}.
By superimposing the electron angular spread on this one-dimensional profile, a full three-dimensional map of Cu–K$_\alpha$ emission can be reconstructed.
At radius $r$, the Cu–K$_\alpha$ emissivity per electron after propagation over a distance $z$ is obtained from
\begin{align}
    j_{\rm 3D}(r,z)=\frac{1}{E_\mathrm{L}}
       \int_{-\infty}^{\infty}
       j_{\rm 1D}\bigl(z,\,T_\mathrm{e}=T_\mathrm{e}(r,t)\bigr)
       \, I(r,t) \dd{t},
\end{align}
with the temporal and spatial weighting supplied by the local laser intensity $I(r,t)$.
Electron-beam divergence was modeled with a normalized Gaussian distribution having a half-angle width $\theta_{\rm dev}$,
\begin{align}
    \mathcal{W}(l,\theta)=N_{0}\exp(-\theta^{2}/\theta_\mathrm{dev}^{2})/l^{2},
\end{align}
where $l$ is the distance from the electron source, $\theta$ is the angle from the laser-propagation direction, and $N_{0}^{-1}=2\pi\int_{0}^{\pi/2}\mathcal{W}(l,\theta)l^2\sin\theta \dd\theta$.
The emission density per steradian at an observation point $(r,z)$ is then
\begin{align}
    \mathbf{J}_{\mathrm{K}_\alpha}(r,z)=
 \iint j_{\rm 3D}\bigl(r', z'\bigr)\,
       \mathcal{W}(z',\theta')\,
       r'\,\dd{r'}\,\dd\theta,
\end{align}
where $z'=\sqrt{\qty(r - r'\cos\theta)^2 + r'^2\sin^2\theta + z^2}$ and $\theta' = \acos(z/z')$ are the distance and polar angle between the electron source and the emission point respectively.
Applying the Beer–Lambert law to account for X-ray absorption inside the tracer and the Abel transform yields the two-dimensional image of Cu–K$_\alpha$ emission:
\begin{align}
    \mathcal{E}_{\rm K_\alpha}(x,z) = \int^{L_{\rm tracer}}_x \frac{r\mathbf{J}_{\mathrm{K}_\alpha}(r,z)}{\sqrt{r^2 - x^2}}\qty{\exp(-\mu\rho \qty(L_{\rm tracer} + r)) + \exp(-\mu\rho \qty(L_{\rm tracer} - r))} \dd{r},
\end{align}
with $2L_{\rm tracer}=500\ \mu$m being the edge length of the tracer, $\mu = 9.2\ \mathrm{cm^2 / g}$ the mass attenuation coefficient that was calculated from PrOpacEOS \cite{MacFarlane2006-hd}, and $\rho = 0.971\ \mathrm{g/cm^{3}}$ the tracer density.
The simulated image was convolved with a two-dimensional Gaussian filter that matches the SCI resolution ($13 \pm 5\ \mu$m FWHM) for quantitative comparison. 
Both the simulated and experimental images were subsequently smoothed with an additional 10-$\mu$m FWHM Gaussian filter to suppress speckle patterns.


\section{Experimental result and discussion}
\label{sec:exp_result}

\begin{figure*}[h]
  \raggedright
  \hspace{0.06\linewidth}
  \textbf{\fontsize{14pt}{14pt}\selectfont (a)} \hspace{0.41\linewidth} \textbf{\fontsize{14pt}{14pt}\selectfont (b)} \\ \vspace{5pt}
  
  \begin{minipage}[h]{0.45\linewidth}
  \centering
  \includegraphics[width=0.8\linewidth]{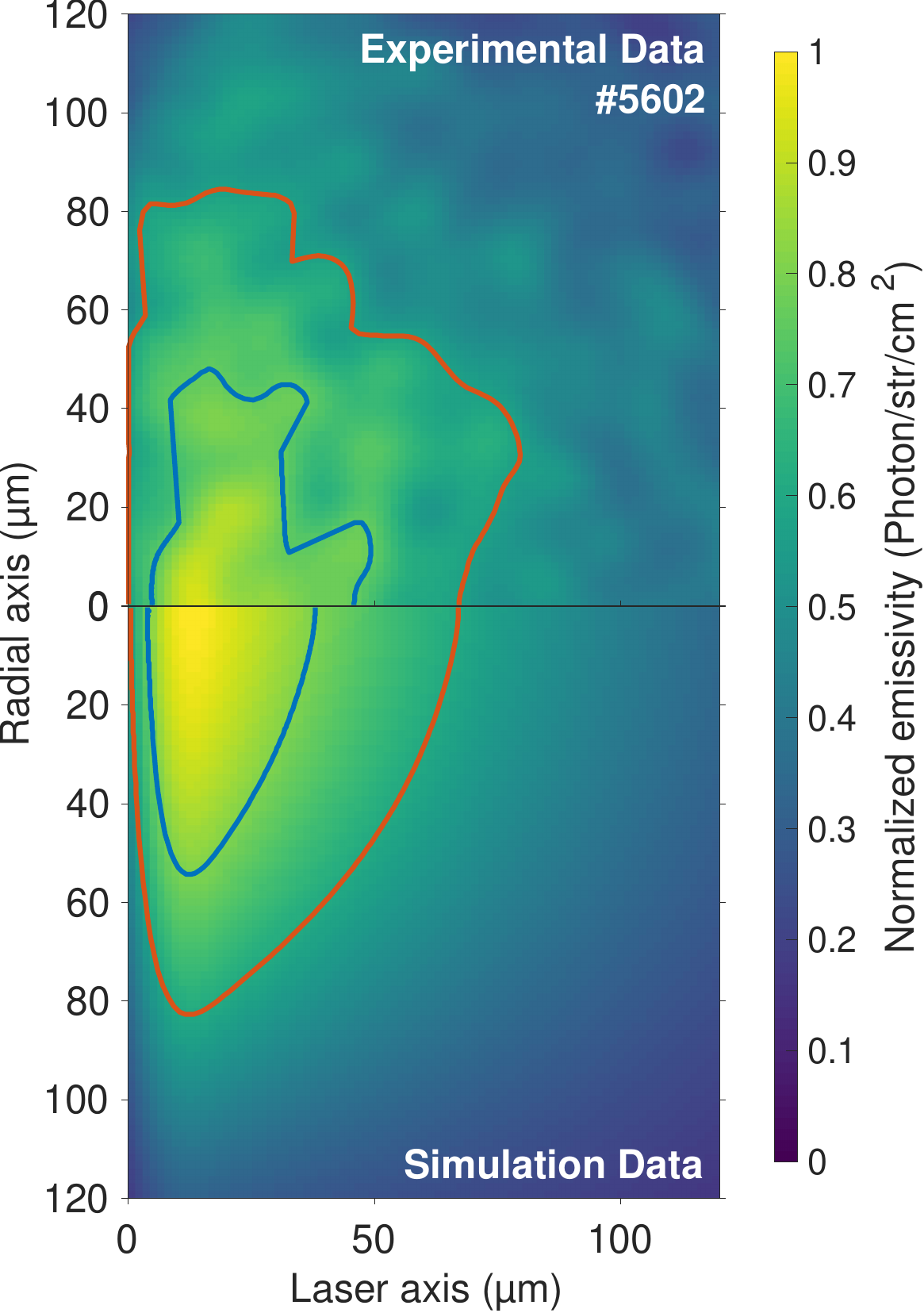}
  \end{minipage}
  \begin{minipage}[h]{0.45\linewidth}
  \centering
  \includegraphics[width=0.8\linewidth]{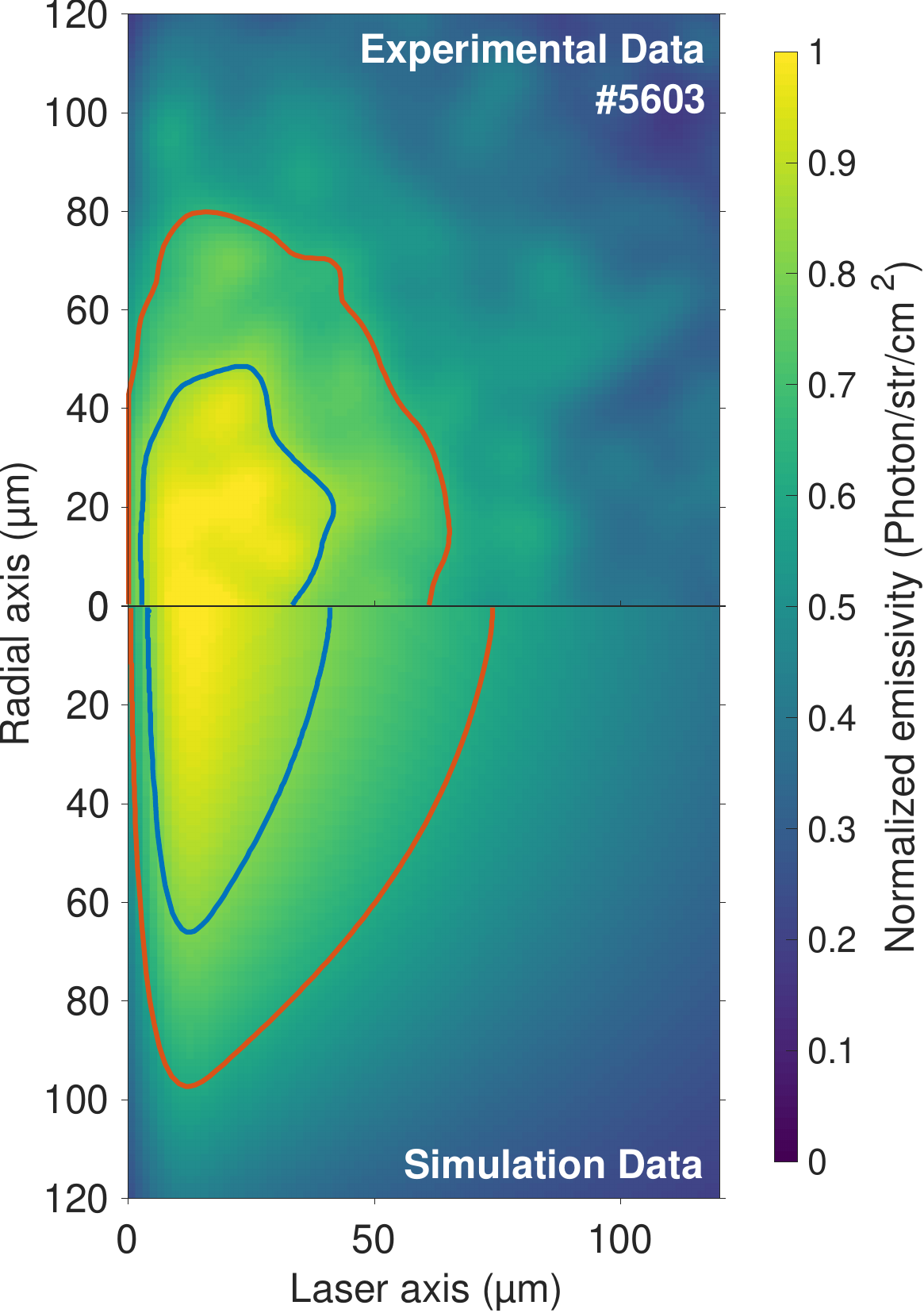}
  \end{minipage}
  \\ \vspace{15pt}
  
  \raggedright
  \hspace{0.06\linewidth}
  \textbf{\fontsize{14pt}{14pt}\selectfont (c)} \hspace{0.41\linewidth} \textbf{\fontsize{14pt}{14pt}\selectfont (d)} \\ \vspace{5pt}
  
  \begin{minipage}[h]{0.45\linewidth}
  \centering
  \includegraphics[width=0.8\linewidth]{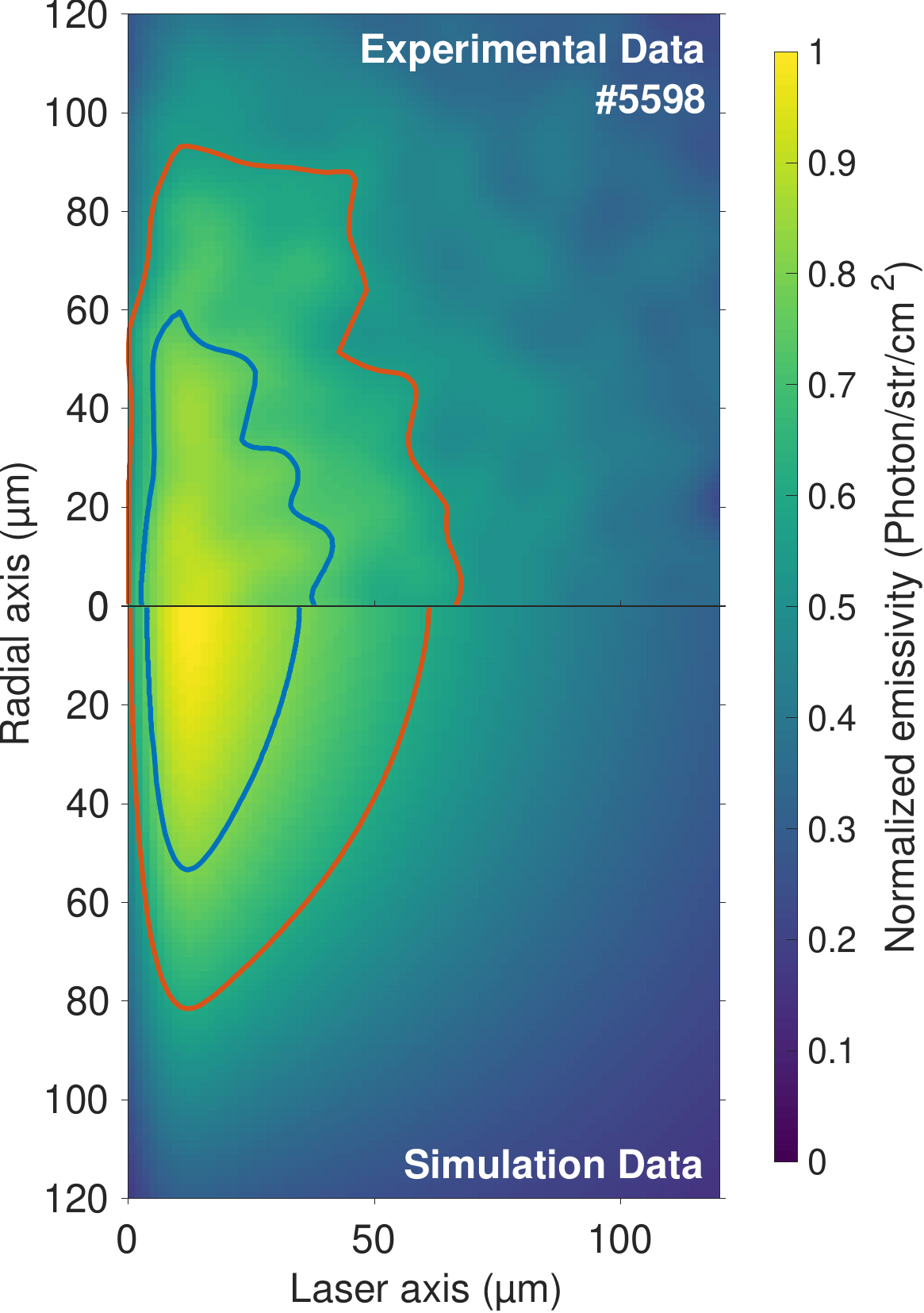}
  \end{minipage}
  \begin{minipage}[h]{0.45\linewidth}
  \centering
  \includegraphics[width=0.8\linewidth]{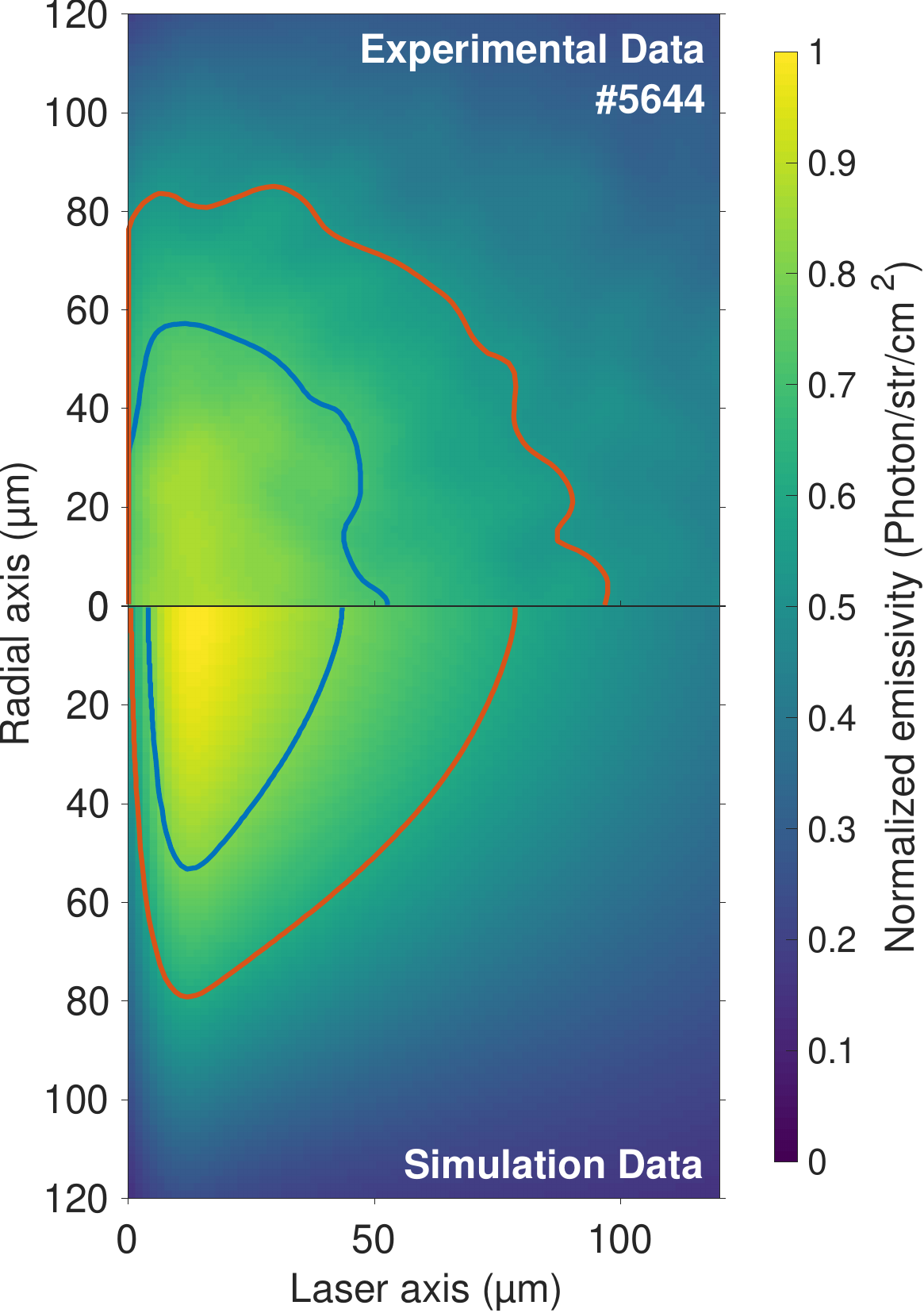}
  \end{minipage}
  
  \caption{
  {\bf Two-dimensional images of Cu–K$_\alpha$ emission.} {\bf (a)} structured target under low-contrast condition, {\bf (b)} planar target under low-contrast condition, {\bf (c)} structured target under high-contrast condition, {\bf (d)} planar target under high-contrast condition.
  The upper and bottom figures represent the experimental and simulation data, respectively.
  }
  \label{fig:comp_image}
\end{figure*}

\begin{figure}[h]
  \centering
  \includegraphics[width=0.8\linewidth]{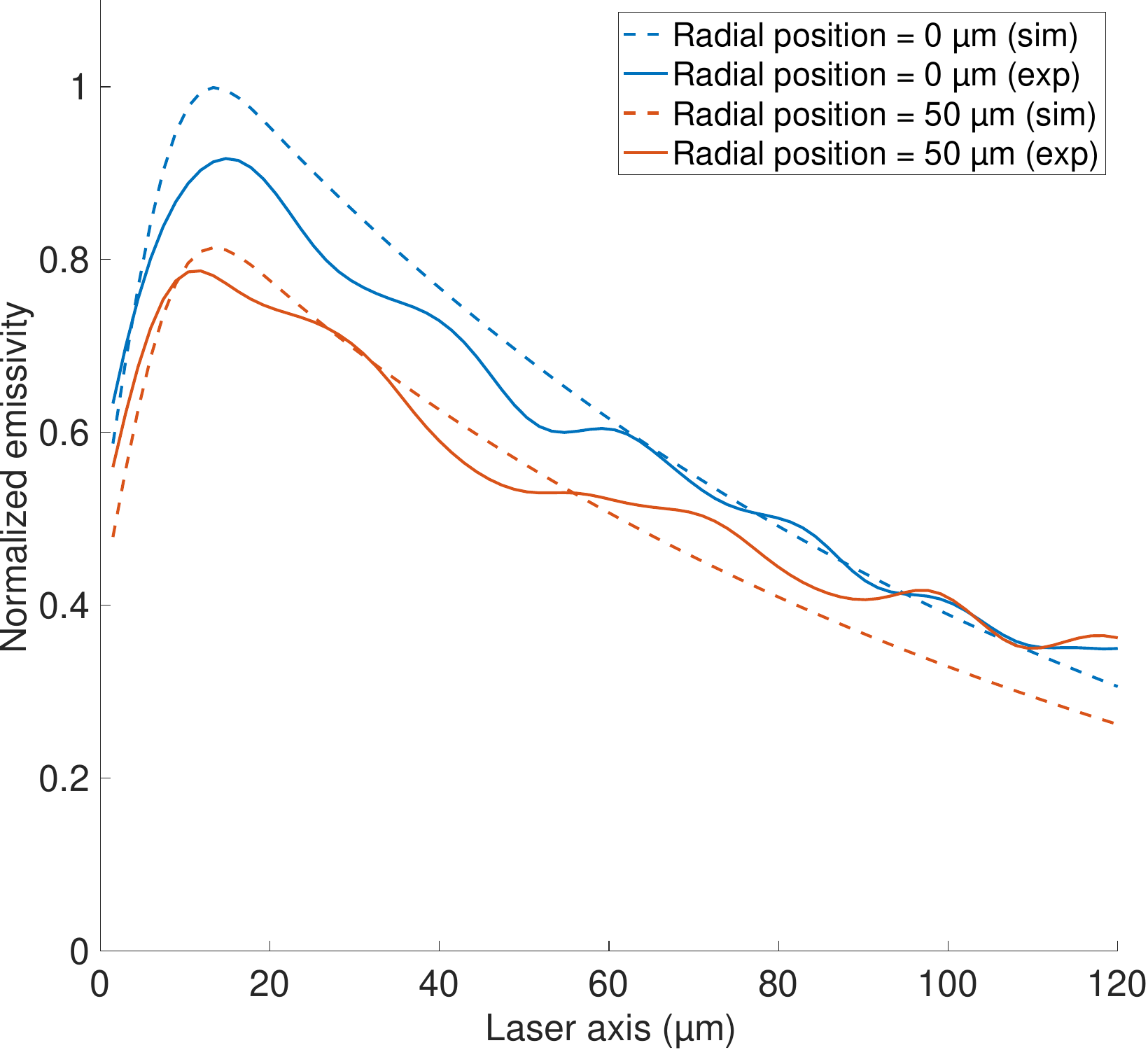}
  \caption{
  {\bf Cu–K$_\alpha$ emission profiles at radial positions 0 $\mu$m and 50 $\mu$m (Shot ID: 5598).} 
  The solid lines denote the experimental measurements, whereas the dashed lines correspond to the simulation results.
  Parameters were optimized by minimizing the difference between the two curves.
  }
  \label{fig:profile}
\end{figure}

\begin{table*}
    \centering
    \scalebox{0.7}{
    \begin{tabular}{|c|c|c|c|c|c|c|}
        \hline
        Shot ID & Contrast & Laser energy (J) & Pulse width (ps) & Spot size ($\mu$m) & Divergence angle (deg.) & Conversion efficiency (\%)  \\ \hline \hline
        \rowcolor{gray} \textbf{5602} & \textbf{Low} & \textbf{330} & \textbf{1.4} & $\bm{240 \pm 30}$ & $\bm{40 \pm 6}$ & $\bm{12 \pm 6}$ \\ \hline
        5603 & Low & 440 & 1.4 & $200 \pm 40$ & $42 \pm 7$ & $14 \pm 6$ \\ \hline
        \rowcolor{gray} \textbf{5598} & \textbf{High} & $\bm{340 \pm 30}$ & \textbf{1.6} & $\bm{235 \pm 35}$ & $\bm{44 \pm 6}$ & $\bm{14 \pm 7}$ \\ \hline
        5644 & High & $350 \pm 30$ & 1.8 & $175 \pm 20$ & $33 \pm 4$ & $4.9 \pm 1.2$ \\ \hline
    \end{tabular}
    }
    \caption{{\bf Characteristics of $J \times B$ accelerated electrons.} The shaded cells show results obtained with the structured targets.}
    \label{tab:res}
\end{table*}

Figure~\ref{fig:comp_image} compares the experimentally measured and numerically simulated two-dimensional images of Cu-K$_\alpha$ emission. 
The origins of the laser and radial axes are the front surface of the tracer and the center of the laser focal spot, respectively. 
The blue and red contours indicate 80\% and 60\% of the peak intensity, and the simulated image reproduces all experimental results well. 
Figure \ref{fig:profile} shows profiles at $r = 0$ and $50~\mu\mathrm{m}$ along the laser-propagation direction; the laser-spot radius and the half-angle scattering width $\theta_{\rm dev}$ are optimized by minimizing the difference between the two curves. 
Table \ref{tab:res} summarizes the derived characteristics of $J \times B$ accelerated electrons. 
Variations in the spot size are not physical effects. 
The failure of the laser-glass cooling system in the laser bay perturbed the beam pointing and enlarged the focus for Shot IDs 5602, 5603, and 5598.

\begin{figure}[h]
  \centering
  \includegraphics[width=0.8\linewidth]{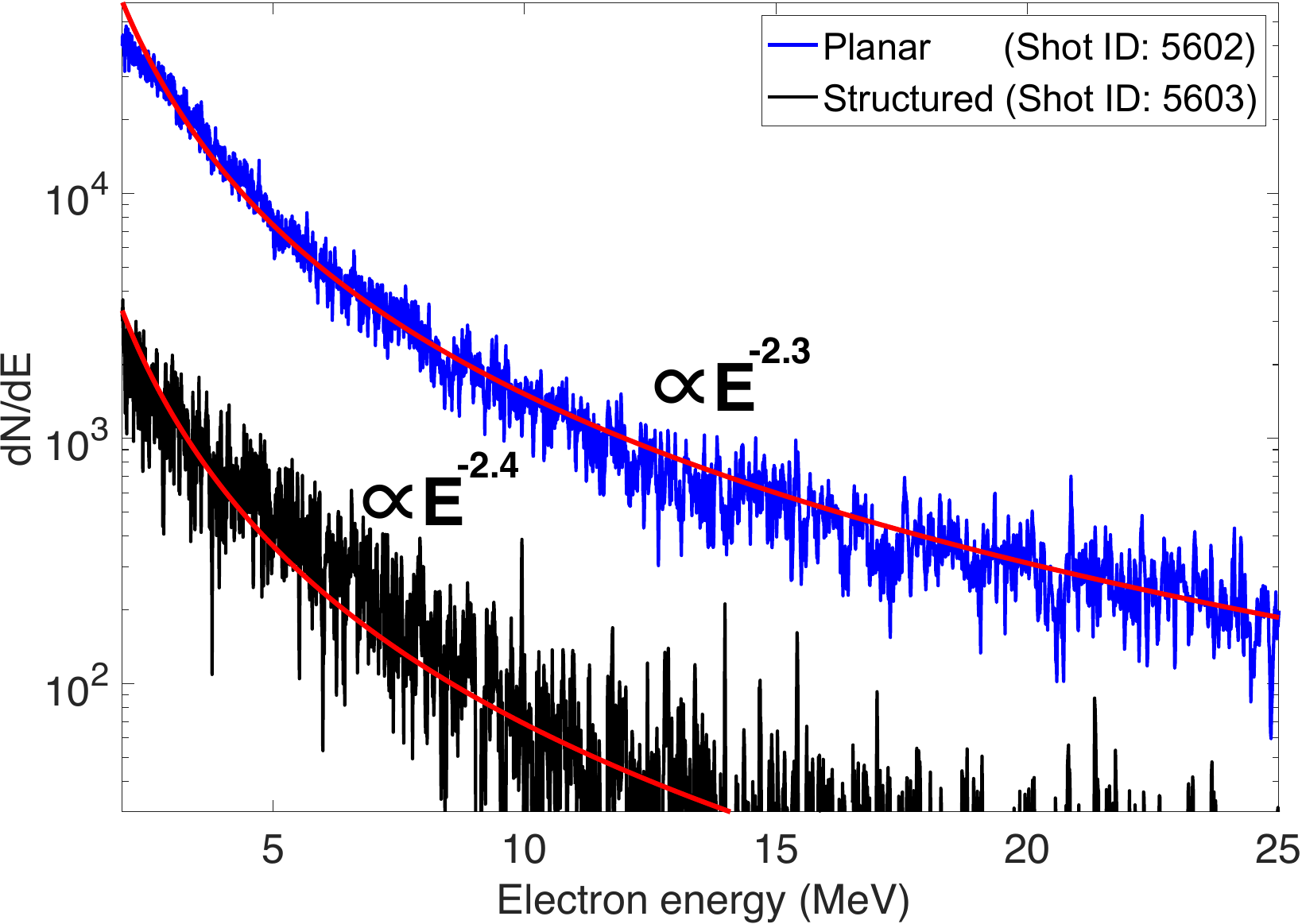}
  \caption{
  {\bf Electron energy distribution under the low-contrast conditions.} 
  Energetic electrons above 2 MeV can penetrate through the 1 mm thickness Ta. 
  Energetic electrons were not detected under the high-contrast conditions.
  }
  \label{fig:ESM}
\end{figure}

Under high-contrast conditions, the structured target nearly tripled the conversion efficiency from laser energy to $J \times B$ accelerated electrons, while electrons whose energies exceeded 2 MeV were still not observed.
This result implies that the enhancement is attributed to an increased number of laser–plasma interaction events rather than to an extended pre-plasma scale length. 
Multiple reflections inside the structure broaden the local incidence angles, which increases the electron divergence.
In the low-contrast case, the structured target does not enhance the conversion efficiency to $J \times B$ accelerated electrons, because the microstructures are completely filled with pre-plasma.
Figure \ref{fig:ESM} presents the measured electron energy distribution. 
The energy distribution follows a power law, indicating that these electrons are generated through stochastic acceleration within the pre-plasma region.
Structured targets markedly reduce the population of electrons produced by stochastic acceleration. 
Because the laser light is reflected at oblique angles rather than along the target normal, the axis of backward-directed acceleration becomes misaligned with the forward-directed acceleration driven by the incident pulse. 
This misalignment disrupts the closed trajectories necessary for an effective stochastic-acceleration cycle.
The population of electrons produced via stochastic acceleration remains small relative to that of electrons accelerated by the $J \times B$ mechanism. 
Although electrons generated by stochastic acceleration possess sufficiently high energy to induce uniform luminescence of the tracer, no such uniform emission is observed.

\section{Conclusions \& future perspectives}

Electrons accelerated by the $J \times B$ mechanism follow ponderomotive scaling, independent of the laser contrast, and thus of the pre-plasma scale length.  
Consequently, the apparent scale-length dependence of electron energies can be understood as a change in the relative contributions of the underlying acceleration mechanisms.
Previous studies inferred plasma heating from electron temperatures obtained from particle diagnostics; however, that inference is flawed.
Electrons whose energies far exceed ponderomotive scaling are generated through stochastic acceleration within the pre-plasma region.
In the absence of pre-plasma, micro-structured targets multiply laser–plasma interaction events, thereby enhancing the conversion efficiency to $J \times B$–accelerated electrons. 
When pre-plasma is present, however, the microstructures become filled, eliminating this advantage and leaving the conversion efficiency unchanged.
We have demonstrated that imparting surface microstructures allows high-contrast laser to achieve conversion efficiency comparable to those obtained with low-contrast laser. 
Nevertheless, the attained efficiencies remain well below the > 50\% benchmark demanded by fast ignition, and significant challenges persist.
Future research should focus on geometries that further multiply interaction events—for example, designs that force the laser pulse to re-enter near-solid-density plasma several times.
 
\section*{Acknowledgements}
The authors thank the technical support staff at The University of Osaka for assistance with the laser operation, target fabrication, plasma diagnostics, and computer simulation.
This work was partially achieved through the use of large-scale computer systems at the Cybermedia Center at The University of Osaka.
This work is supported by the Joint Research Project/Collaboration Research Program between the National Institute for Fusion Science and the Institute of Laser Engineering at The University of Osaka, a Grant-in-Aid for Scientific Research (No. 
25K17369, 
24H00204, 
23K03354, 
23K03360, 
23K20038, 
22H00118 
), JSPS Core-to-Core Program, (Grant No. JPJSCCA20230003), and "Power Laser DX Platform" as research equipment shared in MEXT Project for promoting public utilization of advanced research infrastructure (Program for advanced research equipment platforms) Grant Number JPMXS0450300021.

\section*{AUTHOR DECLARATIONS}

During the preparation of this work the authors used ChatGPT (o3) in order to improve language and readability. 
After using this tool/service, the authors reviewed and edited the content as needed and take full responsibility for the content of the publication.

\bibliography{ref}

\end{document}